\documentstyle[12pt]{article}
\newcommand{\beq}{\begin{equation}}
\newcommand{\eeq}{\end{equation}}
\newcommand{\bea}{\begin{eqnarray}}
\newcommand{\eea}{\end{eqnarray}}

\newcommand{\EW}{electroweak }
\newcommand{\SM}{standard model }
\newcommand{\RG}{renormalization group }
\def\tr{\mathop{\rm tr}}                
\renewcommand{\thefootnote}{\fnsymbol{footnote}}

\def\gtap{\buildrel > \over {_{\sim}}}
\begin{document}
\topmargin=0.0truecm
\oddsidemargin=-0.8truecm
\evensidemargin=-0.8truecm
\setcounter{page}{1}
\begin{flushright}
IC/95/105 \\
\end{flushright}
\vspace*{3.0cm}
\begin{center}
{\bf A DYNAMICAL LEFT--RIGHT SYMMETRY BREAKING MODEL }
\footnote{Talk given at XXXth Rencontres de Moriond ``Electroweak
Interactions and Unified Theories'', Les Arcs, France, March 11-18, 1995.}\\

\vspace*{0.7cm}
{\large E.Kh. Akhmedov} \\
\vspace*{0.5cm}
{\em International Centre for Theoretical Physics,\\
Strada Costiera 11, I-34100, Trieste, Italy \\
and \\
National Research Center ``Kurchatov Institute'',\\
123182 Moscow, Russia }\\
\end{center}
\vspace*{2.5cm}
\begin{center}
ABSTRACT
\end{center}
Left--right symmetry breaking in a model with composite Higgs scalars is
discussed. It is assumed that the low--energy degrees of freedom are just
fermions and gauge bosons and that the Higgs bosons are generated dynamically
through a set of gauge-- and parity--invariant 4-fermion operators. It is
shown that in a model with composite bi-doublet and two triplet scalars there
is no parity breaking at low energies, whereas in the model with  two doublets
instead of two triplets parity is broken automatically regardless of the
choice  of the parameters of the model. For phenomenologically allowed values
of the right--handed scale the tumbling symmetry breaking mechanism is realized
in which parity breaking at a high scale $\mu_R$ propagates down and eventually
causes the \EW symmetry breaking at the scale $\mu_{EW}\sim 100$ GeV.
The model exhibits a number of low and intermediate mass Higgs bosons with
certain relations between their masses. In particular, the $SU(2)_L$ Higgs
doublet $\chi_L$ is a pseudo--Goldstone boson of the accidental (approximate)
$SU(4)$ symmetry of the Higgs potential and therefore is expected to be
relatively light.
\renewcommand{\thefootnote}{\arabic{footnote}}
\setcounter{footnote}{0}
\newpage
Several years ago, a very interesting approach to \EW symmetry breaking was
put forward, so called ``top condensate'' model$^{1)-4)}$. In this model no
fundamental Higgs boson is present;
instead, it is assumed that there is a strong attractive interaction between
the top quarks, which can lead to the formation of the $t\bar{t}$ bound state
playing a role of the Higgs scalar. This interaction is assumed
to result from a new physics at some high--energy scale $\Lambda$, the origin
and precise nature of which are not specified. At low energies this new
physics would manifest itself through non-renormalizable interactions between
usual fermions and gauge bosons. At the energies $E \ll \Lambda$ the lowest
dimensional operators are most important, which are just the four--fermion
(4-f) operators. The simplest gauge--invariant 4-f operator which includes
the heaviest top quark is
\beq
{\cal L}_{4f}=G(\bar{Q}_{Li}t_R)(\bar{t}_R Q_{Li})\,,
\label{4f1}
\eeq
where $Q_L$ is the left--handed doublet of third generation quarks, $G$ is
the dimensionful coupling constant, $G \sim \Lambda^{-2}$, and it is
implied that the colour indices are summed over within each bracket.
This 4-f interaction can be studied analytically in the large $N_c$
(number of colours) limit in the fermion bubble approximation. For large
enough $G$ ($G>G_{cr}=8\pi^2/N_c\Lambda^2$) the \EW symmetry gets spontaneously
broken, $W^\pm{}$ and $Z^0$ bosons and top quark acquire masses, and a
composite Higgs scalar doublet $H\sim \bar{t}_R Q_L$ is formed.

The top condensate approach reproduced correctly the structure of the
low--energy effective Lagrangian of the \SM and demonstrated how the \EW
symmetry breaking can result from a high--energy dynamics. The question
naturally arises as to whether a similar approach can work in more complicated
cases, i.e. whether the Higgs sectors and symmetry breaking patterns of
more complicated models can always be successfully described starting from a
set of gauge--invariant 4-f operators. In this talk I will report the main
results of the analysis of dynamical symmetry breaking in the left--right
symmetric models done in collaboration with M. Lindner, E. Schnapka and
J.W.F. Valle$^{5)}$.

We studied dynamical symmetry breaking in the left--right symmetric (LR)
models based on the gauge group $SU(2)_L \times SU(2)_R \times U(1)_{B-L}$
following the BHL approach to the \SM$^{3)}$.
The Higgs sector of the most popular LR model consists of a
bi-doublet $\phi \sim (2,2,0)$ and two triplets, $\Delta_L \sim (3,1,2)$
and $\Delta_R \sim (1,3,2)$. Assuming that these scalars are composite, their
fermionic content is
$$\phi_{ij} \sim \alpha (\bar{Q}_{Rj}Q_{Li}) + \beta (\tau_2 \bar{Q}_L Q_R
\tau_2)_{ij}+\mbox{leptonic terms}~, $$
\beq
\vec{\Delta}_L \sim (\Psi_L^T C \tau_2 \vec{\tau}\Psi_L),\;\;
\vec{\Delta}_R\sim (\Psi_R^T C  \tau_2 \vec{\tau}\Psi_R)~.
\label{phiij}
\eeq
Here $Q_L,\Psi_L$ ($Q_R,\Psi_R$) are left--handed (right--handed) doublets
of quarks and leptons, respectively; $i$ and $j$ are isospin indices.

In models with Higgs bosons generated by 4-f operators the composite scalars
are, roughly speaking, ``square roots'' of these 4-f operators. One can
therefore obtain the above composite Higgs bosons starting from the 4-f
operators which are ``squares'' of the expressions in eq.~(\ref{phiij}).
A convenient way to study models with composite Higgs bosons is the
auxiliary field technique, in which one introduces the static auxiliary
scalar fields (with appropriate quantum numbers) with Yukawa couplings and
mass terms but no kinetic terms and no quartic couplings.
One can use the equations of motion for these fields to express them in
terms of the fermionic degrees of freedom and recover the initial 4-f
structures.

The static auxiliary fields can acquire gauge invariant kinetic terms and
quartic self--interactions through the radiative corrections and become
physical propagating scalar fields at low energies provided that the
corresponding gap equations are satisfied$^{3)}$. The kinetic terms and
mass corrections can be derived from the 2--point Green function, whereas
the quartic couplings are given by the 4--point functions. Given the Yukawa
couplings of the scalar fields, one can readily calculate these functions in
the fermion bubble approximation, in which they are given by the corresponding
1-fermion--loop diagrams.

It is usually assumed that the Higgs potential of the LR model is exactly
symmetric with respect to the discrete parity symmetry. However, parity can
be spontaneously broken by $\langle \Delta_R \rangle >  \langle \Delta_L
\rangle$ provided $\lambda_2 >
\lambda_1$ where $\lambda_1$ and $\lambda_2$ are the coefficients of the
$[(\Delta_L^\dagger \Delta_L)^2+(\Delta_R^\dagger \Delta_R)^2]$ and
$2(\Delta_L^\dagger \Delta_L)(\Delta_R^\dagger \Delta_R)$ quartic couplings
in the Higgs potential. In the conventional approach,
$\lambda_1$ and $\lambda_2$ are free parameters and one can always choose
$\lambda_2>\lambda_1$. On the contrary, in the composite Higgs approach
based on a certain set of the effective 4-f couplings, the
parameters of the effective Higgs potential are not arbitrary: they are
all calculable in terms of the 4-f couplings $G_a$ and the scale of new
physics $\Lambda$ $^{3)}$. In particular, in the fermion bubble
approximation at one loop level the quartic couplings $\lambda_1$ and
$\lambda_2$ are induced through the Majorana--like Yukawa couplings
$f(\Psi_L^T C \tau_2 \vec{\tau}\vec{\Delta}_L \Psi_L + \Psi_R^T C
\tau_2 \vec{\tau}\vec{\Delta}_R \Psi_R)  + h.c.$.

It is easy to see that to induce the $\lambda_2$ term one needs the
$\Psi_L$--$\Psi_R$ mixing in
the fermion line in the loop, i.e. the lepton Dirac mass term
insertions. However, the Dirac mass terms are generated by the VEVs of the
bi-doublet $\phi$; they are absent at the parity breaking scale which is
supposed to be higher than the \EW scale. Even if parity and \EW symmetry
are broken simultaneously (which is hardly a phenomenologically viable
scenario), this would not save the situation since $\lambda_2$ is finite in
the limit $\Lambda\to \infty$ whereas the diagrams contributing to $\lambda_1$
are logarithmically divergent and so the inequality $\lambda_2>\lambda_1$
cannot be satisfied.

One is therefore led to consider a model with a different composite
Higgs content. The simplest LR model includes two doublets, $\chi_L
\sim (2,1,-1)$ and $\chi_R \sim (1,2,-1)$, instead of the triplets
$\Delta_L$ and $\Delta_R$.
As we shall see, the model with composite doublets will automatically lead
to the correct pattern of the dynamical breaking of parity.
In conventional LR
models, one can have the doublet Higgs bosons without introducing any other
new particles. In our model all scalars are composite, thus
we must introduce additional singlet fermions -- otherwise it would not
be possible to build up the composite $\chi_L$ and $\chi_R$ fields. We
assume that in addition to the usual quark and lepton doublets there is a
gauge--singlet fermion $S_L \sim (1,\;1,\;0)$.
To maintain the discrete parity symmetry one needs a right--handed
counterpart of $S_L$. This can be either another particle, $S_R$, or
the right--handed antiparticle of $S_L$, $(S_L)^c\equiv C\bar{S}_L^T =
S^c_R$. The latter choice is
more economical and, as we shall see, leads to the desired symmetry
breaking pattern. We therefore assume that under parity operation
$S_L \leftrightarrow S^c_R$.
With this new singlet
and usual quark and lepton doublets
one can introduce
the following gauge--invariant 4-f interactions$^5$:
\bea
{\cal L}_{4f}'&=&G_1(\bar{Q}_{Li}Q_{Rj})(\bar{Q}_{Rj}Q_{Li})+
[G_2(\bar{Q}_{Li}Q_{Rj})(\tau_2\bar{Q}_{L}Q_{R}\tau_2)_{ij}
+h.c.]\nonumber \\
& &+G_3(\bar{\Psi}_{Li}\Psi_{Rj})(\bar{\Psi}_{Rj}\Psi_{Li})+
[G_4(\bar{\Psi}_{Li}\Psi_{Rj})(\tau_2\bar{\Psi}_{L}\Psi_{R}\tau_2)_{ij}
+h.c.]\nonumber \\
& &+[G_5(\bar{Q}_{Li}Q_{Rj})(\bar{\Psi}_{Rj}\Psi_{Li})+h.c.]+
[G_6(\bar{Q}_{Li}Q_{Rj})(\tau_2\bar{\Psi}_{L}\Psi_{R}\tau_2)_{ij}
+h.c.] \nonumber \\
& &+G_7[(S_L^T C \Psi_L)(\bar{\Psi}_L C \bar{S}_L^T)+
(\bar{S}_L\Psi_R)(\bar{\Psi}_R S_L)]+G_8 (S_L^T C S_L)(\bar{S}_L C
\bar{S}_L^T)~.
\label{L4f}
\eea
These interactions are not only gauge invariant, but also (for hermitean
$G_2$, $G_4$, $G_5$ and $G_6$) symmetric with respect to the discrete parity
symmetry.

The composite Higgs scalars which can be induced by these 4-f couplings
include, in addition to the bi-doublet $\phi$ of the structure given in
eq.~(\ref{phiij}), two doublets $\chi_L$ and $\chi_R$ and also a singlet
$\sigma$:
\beq
\chi_L \sim S_L^T C \Psi_L, \;\;\;\;  \chi_R \sim \bar{S}_L\Psi_R =
(S^c_R)^T C \Psi_R, \;\;\;\; \sigma \sim \bar{S}_L C \bar{S}_L^T.
\label{composite}
\eeq
Under parity $\chi_L \leftrightarrow \chi_R$, $\sigma \leftrightarrow
\sigma^\dagger$.

In the auxiliary field formalism the scalars $\chi_L$, $\chi_R$, $\phi$ and
$\sigma$ have the following bare mass terms and Yukawa couplings:
\bea
{\cal L}_{aux}&=&-M_0^2(\chi_L^\dagger \chi_L+\chi_R^\dagger \chi_R)
-M_1^2 \tr{(\phi^\dagger \phi)}-\frac{M_2^2}{2}\tr{(\phi^\dagger
\tilde{\phi}+h.c.)}-M_3^2 \sigma^\dagger\sigma \nonumber \\
& & -\left[\bar{Q}_L(Y_1\phi+Y_2\tilde{\phi})Q_R +
\bar{\Psi}_L(Y_3\phi+Y_4\tilde{\phi})\Psi_R + h.c.\right] \nonumber \\
  & &
-\left[
Y_5(\bar{\Psi}_L \chi_L S^c_R+\bar{\Psi}_R \chi_R S_L)
+Y_6 (S_L^T C S_L)\sigma + h.c.
\right]
\label{Laux}
\eea
where the field $\tilde{\phi}\equiv \tau_2\phi^*\tau_2$ has the same
quantum numbers as $\phi$: $\tilde{\phi}\sim (2,\;2,\;0)$.
By integrating out the auxiliary scalar fields one can reproduce the 4-f
structures of eq.~(\ref{L4f}).

Consider now parity breaking in the present LR model.
Using the Yukawa couplings of the doublets $\chi_L$ and $\chi_R$, one can
calculate the fermion-loop
contributions to the  quartic couplings $\lambda_1 [(\chi_L^\dagger
\chi_L)^2+(\chi_R^\dagger \chi_R)^2]$ and $2\lambda_2 (\chi_L^\dagger
\chi_L)(\chi_R^\dagger \chi_R)$ in the effective Higgs potential.
One can easily make sure that the fermion--loop diagrams yield
$\lambda_1=\lambda_2$. Recall that one needs $\lambda_2 > \lambda_1$ to have
spontaneous parity breaking in the LR models. As we shall see, taking into
account the gauge-boson loop contributions to $\lambda_1$ and $\lambda_2$
will automatically secure this relation.

Both $\lambda_1$ and $\lambda_2$ obtain corrections from $U(1)_{B-L}$ gauge
boson loops, whereas only $\lambda_1$ is corrected by diagrams with $W^i_L$
or $W^i_R$ loops.
Since all these contributions have a relative minus sign compared
to the fermion loop ones, one finds $\lambda_2>\lambda_1$ irrespective of
the values of the Yukawa or gauge couplings or any other parameter of the
model, provided that the $SU(2)$ gauge coupling $g_2\neq 0$.
Thus the condition for spontaneous parity breaking is automatically
satisfied in our model.

We have a very interesting situation here. In a model with
composite triplets $\Delta_L$ and $\Delta_R$ parity
is never broken, i.e. the model is not phenomenologically viable. At the
same time, in the model with two composite doublets $\chi_L$ and $\chi_R$
instead of two triplets (which requires introduction of an additional singlet
fermion $S_L$) parity is broken automatically. This means that, unlike in
the conventional LR models, in the composite Higgs approach {\em whether or
not parity is spontaneously broken depends on the particle content of the
model rather than on the choice of the parameters of the Higgs potential}.

Although there are no triplet Higgs bosons in the present version of the
model, a modified seesaw mechanism is operative which ensures the smallness
of the masses of neutrinos taking part in the usual \EW interactions. This
is because the neutrinos mix with the gauge--singlet fermion $S_L$.
Minimization of the effective Higgs potential shows that for the choice of
the 4-f couplings resulting in the dynamical breaking of the LR gauge
symmetry, the $\chi_L$ and $\sigma$ fields do not develop VEVs; for
this reason the lightest neutrinos remain massless in our model.

Analysis of the vacuum structure of the model shows that for
the right--handed scale $v_R=\langle \chi_R^0\rangle$ to lie in the
phenomenologically allowed domain, the effective (mass)$^2$ term of the
composite bi-doublet $\phi$ must always be positive; the \EW symmetry is
broken because of the mixing of $\phi$ with $\chi_R$.
Thus we have a tumbling scenario where the breakdown of parity and $SU(2)_R$
occurring at the scale $\mu_R$ causes the \EW symmetry breaking at a
lower scale $\mu_{EW}$.

The physical Higgs boson sector of the model includes 4 charged scalars,
4 neutral CP--even and 2 CP--odd scalars. Two of neutral CP--even bosons,
$H_1$ and $H_2$, are directly related to the two steps
of symmetry breaking, $SU(2)_R\times U(1)_{B-L}\rightarrow U(1)_Y$ and
$SU(2)_L\times U(1)_Y \rightarrow U(1)_{em}$. In the bubble approximation
the mass of $H_2$, whose properties are similar to the properties of the
\SM Higgs boson, is approximately $2 m_t$. This coincides with the
top--condensate prediction$^{1)-4)}$ and reflects the fact that this scalar
is the $t\bar{t}$ bound state. Analogously, $H_1$ is the bound state of
heavy neutrinos with its mass being approximately twice the heavy neutrino
mass $M \sim \mu_R$.

In conventional LR models only one scalar, which is the analog of the \SM
Higgs boson, is light (at the \EW scale), all the others have their masses
of the order of the right--handed scale $M$. In our
case, the masses of those scalars are also proportional to $M$, but all
of them except the mass of $H_1$ have some suppression factors.
The masses of $\chi_L$ are suppressed because of their
pseudo--Goldstone nature. It was already mentioned that at the
fermion--bubble level $\lambda_1=\lambda_2$ which means that the $(\chi_L,
\chi_R)$ sector of the Higgs potential depends on $\chi_L$ and $\chi_R$
only through the combination $(\chi_L^\dagger\chi_L+\chi_R^\dagger \chi_R)$.
This, in turn, implies that the Higgs potential has a global $SU(4)$
symmetry which is bigger than the original gauge symmetry.
In fact, the origin of this $SU(4)$ symmetry can be traced back to the 4-f
operators of eq.~(\ref{L4f}). It is an accidental symmetry
resulting from the gauge invariance and parity symmetry of the $G_7$ term.
Note that no such symmetry occurs in conventional LR models. After
$\chi_R^0$ acquires a VEV, $SU(4)$ symmetry is broken down to $SU(3)$, and
the components of $\chi_L$ are the corresponding Goldstone bosons. The $SU(4)$
symmetry of the Higgs potential is not exact: it is broken by the $SU(2)$
gauge--boson loop contributions which make $\lambda_2>\lambda_1$ (and also
by $\phi$--dependent terms). As a result, $\chi_L$ are pseudo--Goldstone
bosons with their mass vanishing in the limit $g_2\rightarrow 0$,
$m_{\tau}\to 0$. In fact, though the $SU(2)$ gauge coupling constant $g_2$
is smaller than the typical Yukawa constants in our model, it is not too
small; estimates of the $\chi_L$ mass give $M_{\chi_L} \sim 10^{-1} M$.

The bi-doublet $\phi$ can be viewed as consisting of two doublets, $\phi_1$
which develops a VEV and is similar to the \SM Higgs doublet, and the
orthogonal field $\phi_2$ which is VEVless. In the conventional LR models
$\phi_2$ is heavy, $M_{\phi_2}\sim M$. In our case the mass of its charged
components  $\phi_2^\pm$ is suppressed by the factor $m_{\tau}/m_t$ and
is therefore of the order $10^{-2} M$. The masses of the neutral CP--even
and CP--odd components are even smaller; they are related to the masses of
$\phi_2^\pm$ and the \SM Higgs boson $H_2$ by
\beq
M_{\phi_{2r}^0}^2=M_{\phi_{2i}^0}^2=M_{\phi_2^\pm}^2-\frac{M_{H_2}^2}{2}
=\frac{2}{3}M^2\frac{m_{\tau}^2}{m_t^2}-\frac{M_{H_2}^2}{2}.
\eeq
This equation imposes an upper limit on the
\SM Higgs boson mass $M_{H_2}$ (for a given $M$) or a lower limit on the
right--handed mass $M$ (for a given $M_{H_2}$). These limits follow from the
requirement that $M_{\phi_2^\pm}^2$ be positive, i.e. from the vacuum
stability condition. For example, for $M_{H_2} \approx 200$ GeV we find
$M\gtap 17$ TeV.

Assuming that only one of the neutral components of the bi-doublet develops
a non-vanishing VEV, one can readily relate the top quark mass to the scale
of new physics $\Lambda$ (on which it depends logarithmically) and the \EW
VEV. For example, for $\Lambda\simeq 10^{15}$ GeV, one finds $m_t\simeq 165$
GeV. However, this only holds in the bubble approximation; the \RG improved
result is substantially higher, 220--230 GeV. In this respect the predictions
of the model are similar to those of BHL$^{3)}$. If one assumes
that both neutral components of the bi-doublet develop non-vanishing VEVs,
one can easily get an acceptable value of $m_t$. However, the top quark mass
is adjusted rather than predicted, i.e. one looses predictivity in the
fermion sector in this case (although keeps interesting predictions in the
Higgs boson sector).

To summarize, we have a successful dynamical LR model in which a tumbling
symmetry breaking mechanism is operative. The model exhibits
a number of low and intermediate scale Higgs bosons and predicts the
relations between masses of various scalars and
between fermion and Higgs boson masses which are in principle testable.
If the right--handed scale $\mu_R$  is of the order of a few tens of TeV,
the neutral $CP$--even and $CP$--odd scalars $\phi_{2r}^0$ and $\phi_{2i}^0$
can be even lighter than the \EW Higgs boson. In fact, they can be as light
as $\sim 50$ GeV and so might be observable at LEP2. Such light $\phi_{2r}^0$
and $\phi_{2i}^0$ can also provide a positive contribution to
$R_b=\Gamma(Z\to b\bar{b})/\Gamma(Z\to hadrons)$ which is necessary to
account for the discrepancy between the LEP observations and the \SM
predictions. \\

\centerline{REFERENCES}
\vspace*{0.2cm}
\baselineskip=.3truecm
\noindent
1. Y. Nambu, in {\em New Theories in Physics, Proc. XI Int. Symposium on
Elementary Particle \hspace*{0.4truecm} Physics}, eds. Z. Ajduk, S. Pokorski
and A. Trautman (World Scientific, Singapore, 1989) \hspace*{0.4truecm} and
EFI report No. 89-08 (1989), unpublished.  \\
2. A. Miransky, M. Tanabashi, K. Yamawaki, Mod. Phys. Lett. A4 (1989) 1043;
Phys. Lett. \hspace*{0.4truecm} B221 (1989) 177. \\
3. W.A. Bardeen, C.T. Hill, M. Lindner, Phys. Rev. D41 (1990) 1647.  \\
4. W.J. Marciano, Phys. Rev. Lett. 62 (1989) 2793. \\
5. E.Kh. Akhmedov, M. Lindner, E. Schnapka, J.W.F. Valle, to be published.\\
\end{document}